\documentclass[12pt,preprint]{aastex}
\usepackage{epsfig,emulateapj5,apjfonts,aastexug,onecolfloat}

\def\gtsima{$\; \buildrel > \over \sim \;$}
\def\ltsima{$\; \buildrel < \over \sim \;$}
\def\prosima{$\; \buildrel \propto \over \sim \;$}
\def\gsim{\lower.5ex\hbox{\gtsima}}
\def\lsim{\lower.5ex\hbox{\ltsima}}
\def\simgt{\lower.5ex\hbox{\gtsima}}
\def\simlt{\lower.5ex\hbox{\ltsima}}
\def\simpr{\lower.5ex\hbox{\prosima}}

\def\h1{$h^{-1}$}
\def\eeq{\end{equation}}
\def\beq{\begin{equation}}

\submitted{Submitted ; Accepted}

\shorttitle{Clustering of red high-$z$ galaxies}
\shortauthors{E. Daddi et al.}

\journalinfo{ApJ accepted}
\begin{document}

\twocolumn[

\title{Detection of strong clustering of red $K$-selected galaxies\\ at $2<\lowercase{z_{phot}}<4$ in the Hubble Deep Field South 
\footnotemark[1] $^,$
\footnotemark[2]
}

\author{E. Daddi\altaffilmark{3}, 
	H.J.A. R\"ottgering\altaffilmark{4},
	I. Labb\'e\altaffilmark{4},
	G. Rudnick\altaffilmark{5},
	M. Franx\altaffilmark{4},
	A.F.M. Moorwood\altaffilmark{3},\\
	H.W. Rix\altaffilmark{6},
	P.P. van der Werf\altaffilmark{4},
	P.G. van Dokkum\altaffilmark{7,8}
}
\affil{$^{3}$European Southern Observatory, Karl-Schwarzschild-Str. 2, D-85748
Garching, Germany \email{edaddi@eso.org}}
\affil{$^{4}$Sterrewacht Leiden, Postbus 9513, 2300 RA Leiden, The Netherlands}
\affil{$^{5}$Max-Plank-Institut f\"ur Astrophysik, Karl-Schwarschild-Str. 1, Garching, Germany}
\affil{$^{6}$Max-Plank-Institut f\"ur Astronomie, K\"onigstuhl 17, Heidelberg, Germany}
\affil{$^{7}$California Institute of Technology, 1200 East California Boulevard, Pasadena (CA), USA}
\affil{$^{8}$ Present address: Department of Astronomy, Yale University, New Haven, CT 06520-8101}

\begin{abstract}
The clustering properties of faint $K_{VEGA}<24$ galaxies
are measured in ultradeep J, H and K near-IR images of the Hubble Deep Field 
South (HDF-S), obtained with ISAAC at the VLT. 
As a function of the $K$-magnitude, a relatively large clustering amplitude is
found up to $K=24$, at a level comparable to the measurements at $K\sim19$.
The photometric redshift distribution of $K<24$ galaxies extends
to $z_{phot}\sim4$--5, and $\sim40$\% of the galaxies are at $z_{phot}>2$. 
At the highest redshifts, $2<z_{phot}<4$, galaxies selected in the rest 
frame optical ($K$-band)
appear significantly more clustered than galaxy populations selected in the rest
frame UV (i.e. Lyman Break Galaxies, LBGs), in a similar redshift range
and with similar number densities.
Galaxy clustering depends on the $J-K$ color at $2 < z_{phot} < 4$, 
with the $K$-selected galaxies with $J-K>1.7$ reaching $r_0\sim8$ \h1 Mpc
comoving. This is a factor of 3--4 higher than the correlation length of LBGs
with similar number densities, down to $V_{606}<27$, and is also larger than
the correlation length of $K$-selected galaxies with blue $J-K<1.7$ colors.
Hence at $z\approx3$ a  color-density relation is observed which
is qualitatively  similar to that observed locally. Fluctuations in the 
amplitude of clustering due to cosmic variance may affect our estimates derived
from the small HDF-S field, but these are unlikely to change our main 
conclusions.
The galaxies with red $J-K>1.7$ colors at $2<z_{phot}<4$ 
are likely older and more massive galaxies, on average, than LBGs. They were 
presumably formed in the highest density perturbations at early epochs.
Semi-analytical hierarchical models do predict the existence of
strongly clustered populations at $z\sim3$, but with at least a factor 
of 10 lower number density than the one measured.
The overall properties of this strongly clustered 
population consistently
suggest that they are the progenitors, or building blocks,
 of local massive early-type galaxies
and $z\sim1$ EROs, close to their major epochs of formation.
\end{abstract}
]

\section{Introduction}

\footnotetext[1]{
Based on observations collected  
at the European Southern Observatory, Paranal, Chile (ESO LP 164.O-0612).}
\footnotetext[2]{Based on observations with the NASA/ESA {\em Hubble Space Telescope}, 
obtained at the Space Telescope Science Institute, which is operated by AURA 
Inc, under NASA contract NAS 5-26555.}
Measurements of clustering at large redshifts can be used to shed 
light on the assembly of large scale structure in the Universe and
to trace the history and evolution of galaxies. 

Locally the Sloan Digital Sky Survey (Stoughton et al. 2002) 
and the "2 degree field Galaxy Redshift Survey" ({\em 2dfGRS}, Colless et al. 2001)
are now providing a detailed description of galaxy clustering
(Norberg et al.  2002; Zehavi et al. 2002). 
Redshift surveys in the last few years were also able to provide a first
glimpse of the evolution of galaxy clustering up to $z\sim1$ by direct
real space measurements
(LeF\`evre et al.  1996; Carlberg et al.  1997; Hogg et al.  2000), 
finding a general decrease of the clustering strength with redshift. 
At higher redshift, significant clustering of Lyman
Break Galaxies (LBGs) at $z\sim3$ (Giavalisco et al.  1998; Adelberger et al.
1998) has been detected, 
implying a large bias for this population of high-$z$ galaxies,
in qualitative agreement with hierarchical scenarios (e.g. Baugh et
al.  1998). 
Ouchi et al.  (2001; 2003) 
recently confirmed these high bias values on the basis of an extensive
survey for LBGs 
at $z\sim4$ and for Ly$\alpha$ emitting galaxies at $z=4.86$, in the
Subaru/{\em{XMM}} Deep Survey field.
Giavalisco \& Dickinson (2001, GD01 hereafter)
presented evidence of luminosity segregation in the clustering of LBGs, with
the brighter LBGs being more strongly clustered than the fainter ones, possibly
consistent with the expectations of hierarchical clustering.

A good approach to address the clustering of high redshift galaxies,
where spectroscopic redshifts are hard or impossible to obtain, is
to use photometric redshifts to identify galaxy populations from fields
with deep photometric data. As compared to the pure LBG selection, this
offers the advantage of having, in principle, less of a selection bias
against reddened or weakly star-forming galaxies, and of allowing access to
a wider redshift range.
The most useful fields at this point are the Hubble Deep Fields (HDFs), owing
to the very deep and accurate photometry available. Magliocchetti \&
Maddox (1999) and Arnouts et al. (1999; 2002) present clustering
measurements of $I$-band selected HDF-S galaxies extending up to
$z_{phot}\sim4$--5, suggesting a trend of increasing clustering for
$z_{phot}\simgt2$ consistent with a positive evolution in the bias.

Despite an overall agreement between the observations and theory, 
a number of issues remain and deserve attention.
Locally, it is well established that galaxies of different types 
cluster very differently (e.g. Guzzo et al. 1997). At $z\sim1$ some
evidence of a significantly different clustering of early-type and
late-type galaxies is beginning to emerge as well (e.g. Carlberg et al. 1997; 
Daddi et al. 2002).
Since at higher redshift the clustering strength of various
classes of objects has not yet been measured,
it is difficult to reconcile the globally measured clustering at high 
$z$ with the detailed local measurements of clustering.
Furthermore, it is still controversial what  the present-day
descendants of LBGs are and so it is difficult to place them in the
context of a global evolutionary scenario of clustering.
Similarly, the decreasing trend of clustering amplitude measured to
$z\sim1$ in redshift surveys is most likely due to the change of the
galaxy mix with redshift in magnitude selected samples
and may not reflect physical evolution. 

In this paper, we present the  clustering measurement of faint
$K$-selected galaxies in the HDF-S images, obtained for the 
Faint InfraRed Extragalactic Survey (FIRES), and
based on very deep
near-IR observations (more than 33 hours each for the $J$, $H$ and $K$ bands\footnote{The filters used for the observations were the $Js$ and $Ks$ but are
referred to as $J$ and $K$ in the rest of the paper})
carried out with ISAAC at the VLT (Labb\'e et al. 2003).
The FIRES survey (Franx et al. 2000)
was designed to study the evolutionary properties of
galaxies selected in the restframe $\lambda>5000$ \AA\ up to $z\sim3$--4,
with a selection much closer to a selection by stellar mass than 
by optical/UV light. The clustering properties of
galaxies in the MS1054-03 FIRES field (Forster-Schreiber et al. 
in preparation) will
be discussed in a forthcoming paper.
The resulting near-IR selected sample of HDF-S galaxies, 
augmented with the high
quality HST multiband optical photometry, allows a study of 
the evolution of galaxy clustering that is complementary 
to the previous ones based on optical selection.

The paper is organized as follow: the data and measurement techniques are
described in Sect.~\ref{sec:data}; Sect.~\ref{sec:clustering} describes
the angular and spatial clustering estimates, for the whole sample and 
as a function of redshift and colors, focusing in particular on the
clustering properties of $I-K$ and $J-K$ red galaxies.
A comparison with relevant literature data is also presented. 
We analyze the implications of our
findings with some modeling in Sect.~\ref{sec:models}, where we discuss
the theoretical implications of our detection of a strongly clustered
population of $z\sim3$ galaxies.
The summary and conclusions of this work are in Sect.~\ref{sec:concl}.
We assume $\Omega_\Lambda = 0.7$, $\Omega_m = 0.3$ and
$H_0 = 100h$ km/s/Mpc throughout the paper. 
Magnitudes are given on the Vega scale and
distances are expressed in comoving units throughout the paper 
(unless otherwise explicitly stated).

\section{Data and techniques}
\label{sec:data}

\subsection{The catalog and photometric redshifts}

The reduction, identification of galaxies and the photometric measurements, 
as carried out in the context of the FIRES project, are
described in detail by Labb\'e et al.  (2003). For the present paper
a subsample of K-selected galaxies is derived from the Labb\'e et al.
(2003) catalog\footnote{available at http://www.strw.leidenuniv.nl/$\sim$fires/data/hdfs/}, drawn from the $\sim4.5$ arcmin$^2$ ISAAC area containing 
$\geq 95$\% of the total exposure time
in the K-band (more than $10^5$s). 
This selection results in a sample whose depth is uniform.
{The subsample contains 435 galaxies to the completeness limit of
$K=24$, determined via the $K$ band turnover,
with $J$- and $H$- band photometry available.  The seeing of the reduced
near-IR mosaics is about $0.45\arcsec$\ for the J, H and K images.
Most of this area
(i.e. $\sim 4$ arcmin$^2$) is covered by deep HST images  
in the $U_{300}$, $B_{450}$, $V_{606}$, $I_{814}$ bands (Casertano et
al.  2000). The $1\sigma$ sky noise limiting magnitudes over $0.7\arcsec$\
apertures are 29.5, 30.3, 30.6, 30.0, 28.6, 28.1, 28.1 for $U_{300}$,
$B_{450}$, $V_{606}$, $I_{814}$, J, H, K, respectively (all given on the AB
magnitude system, Labb\'e et al. 2003). 
The very deep and high quality imaging available
permits an accurate photometric redshifts estimate for the sample.
Our implementation of the photometric redshift measurements 
is described in detail in Rudnick et al. (2001; 2002 in preparation)
and Labb\'e et al. (2003). Good agreement is found between the
photometric and spectroscopic redshifts for the objects with known
spectroscopic redshift (Vanzella et al. 2002 and reference therein),
with an overall $\Delta z/(1+z_{spec})\approx0.09$ over the whole range,
and even better ($\approx0.05$) at $z>2$.}
The spectroscopic redshift is adopted when available, 
i.e. for about 10\% of the sample.

\subsection{Angular and real space clustering}

A standard approach, based on the minimum variance 
Landy \& Szalay (1993) estimator, is
used to measure the two point correlation function of galaxies
$w(\theta)$. For a detailed discussion of the steps involved we refer
to Daddi et al.  (2000b) and references therein. The 
power-law slopes of the correlation functions are fixed to the standard 
value of $\delta=0.8$. This implies a slope of $\gamma=1.8$ for the 
real space two point correlation function $\xi(r)$. 
We generate random samples with typically 100 times or more objects
than in the observed sample. 
{Following the results of Monte Carlo simulations of Daddi et al. (2001),
we used as errors for the $w(\theta)$ measurements in each bin
$\sigma_w=[(1+w)/DD]^{1/2}$, where DD is the number of observed pairs.
Note that this error only takes into account effects due to the finite number
of objects in the sample. However, due to cosmic variance, it is possible
that there are significant variations of $w(\theta)$ from field to field.
To address how representative the present measurements are for assessing
the average signal, we used the analysis of
Daddi et al.  2001 (see also Bernstein 1994, Arnouts et al.  2002): 

\beq
\sigma(A) = A^{3/2}C^{1/2}
\label{eq:sn}
\eeq

\begin{figure}[ht]
\centerline{\epsfig{file=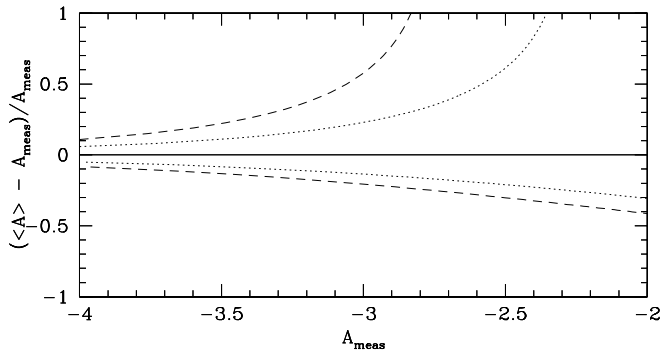,width=9cm}}
\caption{
{The figure describes the relative fluctuations of the clustering 
amplitude expected in a 4 arcmin$^2$ field as due to cosmic variance.
The {\em average} amplitudes of angular clusterinag ($<A>$) that could 
have produced the measured amplitude ($A_{meas}$)
as a result of a 
fluctuation due to cosmic variance can be derived from the dotted 
and dashed lines (1 and 3 sigma levels respectively). 
As can be seen, the curves are highly asymmetric around the
measured value, generally disfavoring the possibility that the {\em average}
level of clustering can be strongly overestimated as due to the cosmic 
variance.}
}
\label{fig:sn}
\end{figure}
where $C$ is a function of the field geometry
($C\propto Area^{-0.4}$). 
This relation suggest that, when the clustering of a galaxy population 
is intrinsically large, large
fluctuations from cosmic variance are expected for the measured
amplitudes (Fig.~\ref{fig:sn}), and relatively small clustering 
amplitudes can be obtained by chance.  On the other hand, if
the clustering is intrinsically small also small fluctuations are
expected. Measures of relatively large clustering amplitudes
are therefore unlikely to arise due to cosmic variance fluctuations from
a population with small average clustering.}

\begin{figure}[h]
\centering 
\includegraphics[width=9.0cm]{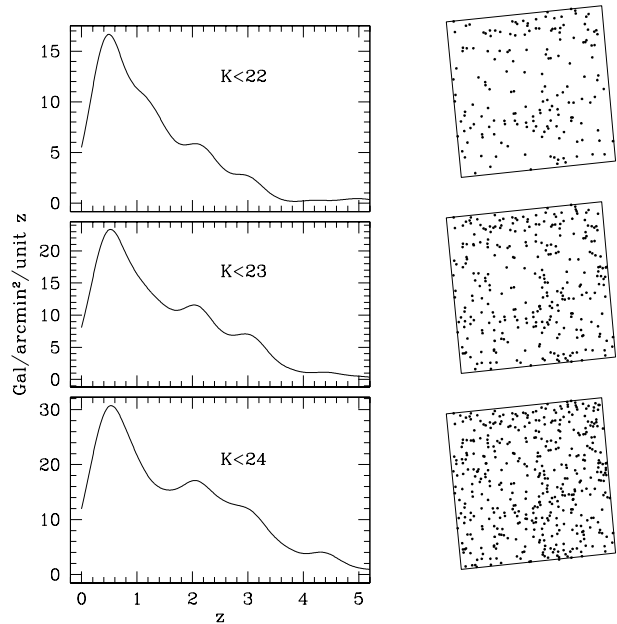}
\caption{
The sky distribution (right panels) and redshift distributions (left
panels, smoothed with a Gaussian with $\sigma_z\sim0.25$) 
for galaxies to the limiting magnitudes $K=22,23,24$. 
}
\label{fig:distri}
\end{figure}

\begin{figure*}[ht]
\centering 
\includegraphics[width=6.7cm,angle=-90]{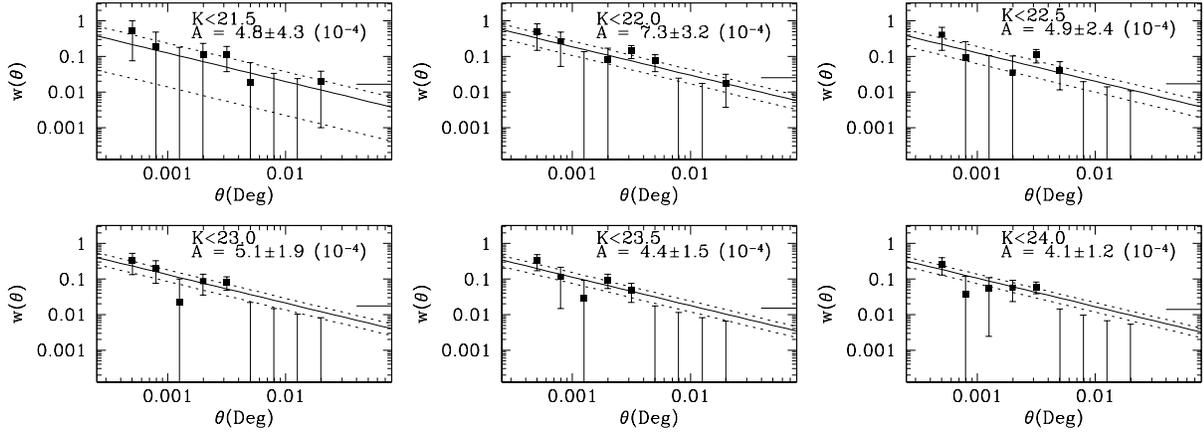}
\caption{
Angular two point correlation function measurements 
for the galaxies for various $K$ limiting
magnitudes. We show the best fit power law $w(\theta) = A\ \theta^{-0.8}$
(solid line)  and the allowed $1\sigma$ range (dotted lines).
}
\label{fig:cors}
\end{figure*}

\begin{figure}[ht]
\includegraphics[width=9cm]{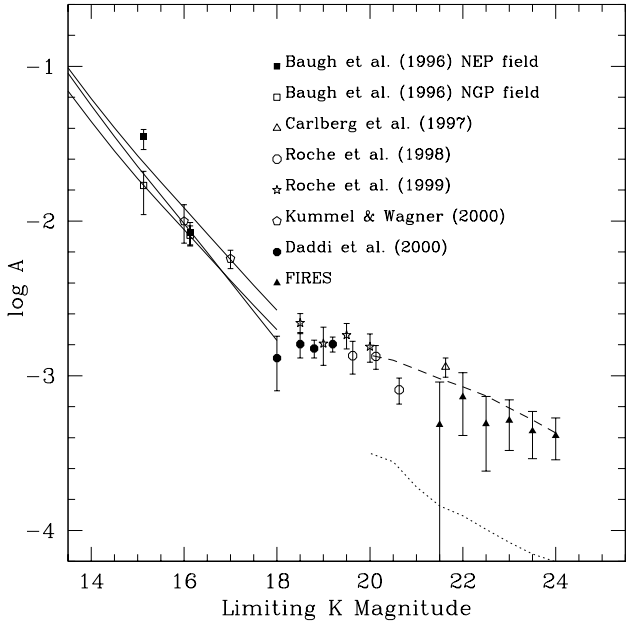}
\caption{Plot of the measured angular clustering amplitudes in the FIRES
survey together with measurements taken from the literature. 
At bright magnitudes we show PLE models from Roche et al.  (1998) with
scaling from local galaxy clustering. At $20<K<24$ we show predictions
for the clustering of $K$-selected galaxies based on (dotted line) an
evolving model that can reproduce the observed $I$-band clustering at
faint magnitudes and (dashed line) the same with enhanced clustering for
early-type galaxies with $r_0=10$ \h1 Mpc at all redshifts (see Sect.~\ref{sec:models} for more details).
}
\label{fig:roche}
\end{figure}

From a given angular clustering measurement, we use the Limber equation to 
infer the real space clustering ($r_0$) in comoving units 
and the {\em effective} redshift 
($\overline{z}$, defined as the redshift for which $r_0$ has been
estimated, i.e. $r_0\equiv r_0(\overline{z})$,
see Daddi et al. 2001 for more details). 
This requires the use of the redshift
selection function for the examined sample, which we derive from the
photometric redshift distribution, appropriately convolved with an error
distribution function for the photometric redshifts. 
For this we typically used Gaussian functions with $\sigma_z=0.25$ in
order to wash out structures in the observed photometric distributions.
Variations around this value (e.g. from $\sigma_z=0.1$ to
0.5) produce only minimal variations in the estimated $r_0$, well below
the quoted uncertainties.

{A subsample selected for a certain range in photometric redshifts
will also contain a certain fraction of galaxies with true redshift 
outside this range. In general these galaxies would dilute the clustering
signal. The measured signals can therefore be regarded as lower limits
to the true signal, in those cases.
}

\section{Results on the angular and real space amplitude}
\label{sec:clustering}

In this section we will present the  results of our study of the clustering
of galaxies in the HDF-S as a function of magnitude, redshift and color.
Table~\ref{tab:all} summarizes all
the angular clustering measurements 
presented in the paper together with the relevant derived quantities.

\subsection{Clustering of galaxies to $K=24$}

Fig.~\ref{fig:distri} shows the sky and redshift distributions for
the galaxies in our survey. The redshift distribution of $K\leq24$
galaxies peaks at $z_{phot}\simlt1$ and has a tail extending to $z_{phot}\sim5$,
becoming more and more prominent at fainter magnitudes. Its overall
shape at $K\leq24$ is similar to that derived from the Subaru Deep Field
(Kashikawa et al. 2003).
An inhomogeneous angular distribution of galaxies is apparent from these plots. 
Measurements of the angular two point 
correlation functions are shown in Fig.~\ref{fig:cors}: the clustering 
of galaxies is detectable down to the faintest levels with $S/N\sim3$.
The overall amplitude of clustering in all cases is low enough ($A<10^{-3}$, 
Fig.~\ref{fig:cors}) to allow us to
neglect the effects of cosmic variance (Fig.~\ref{fig:sn}). 
These are the first measurements of clustering to date extending to magnitudes
$K\geq 22$. 

In Fig.~\ref{fig:roche} our measurements are compared with a collection of
literature data at brighter magnitudes. 
The new FIRES measurements clearly indicate that the amplitude of clustering
remains relatively flat, 
at the level of the $K\sim19$ measurements, all the way to
$K=24$ with only a mild decline. 
Preliminary indications of this behavior were already found at $K=21.5$ by
Carlberg et al.  (1997, a result also discussed by Roche et al.  1998), 
with a measurement consistent with 
our own at this bright $K$ level.
This trend is significantly different from the most recent
measurements in optical imaging surveys. In the $R$- and $I$-bands
a monotonically decreasing clustering amplitude is
observed to the faintest magnitudes of $R\sim29$ and $I\sim28$
(Postman et al.  1998, Villumsen et al.  1997, McCracken et al.  2001). 
The FIRES galaxies with $K\leq 24$ have a median color of $I-K\sim2.4$ 
and a median $I\sim25$ magnitude\footnote{Our $I$-band measurements refer
to the $I_{814}$ filter magnitude, expressed in the Vega scale. A color term
from the standard Cousins $I$-band may be expected of order of 0.1
magnitudes}. McCracken et al.  (2001) measure, for
galaxies with median $I\sim25$, an angular clustering that is 3-4 times
lower than the one of FIRES galaxies.

Inverting the angular clustering with the Limber equation, 
using the redshift distributions given in Fig.~\ref{fig:distri}, we
find for $K\leq24$ that the observed level of clustering requires 
$r_0=3.5$--4.5 Mpc at 
an {\em effective} redshift of $z=1.3$. 
This value is similar to
the typical correlation length of the faintest {\it 2dfGRS} 
galaxies at $z=0$ (Norberg et al. 2002). 
Because they are intrinsically fainter, 
the galaxies with $z_{phot}\simlt 1$ are expected to
have a similar or even lower correlation length.
The $z_{phot}>1$ galaxies, for consistency,
should also have a similar or higher correlation length,
up to $z_{phot}\simgt3$--4 where the tail of the galaxy distribution extends,
to produce the measured angular correlations.

These results show clearly that faint $K$-selected samples are intrinsically
more clustered than optically selected samples. The interpretation for
the clustering trend at very faint $K$ levels will be explored further in
Sect.~\ref{sec:models} with some modeling.

\subsection{Clustering as a function of redshift}

\begin{figure}[ht]
\includegraphics[width=9cm]{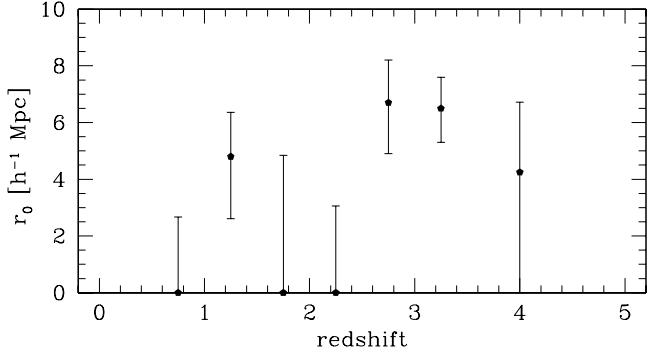}
\caption{
The correlation length of $K<24$ galaxy samples as a
function of redshift.
}
\label{fig:zev}
\end{figure}

The galaxies with $K<24$ were divided in photometric redshift bins 
in order to measure the redshift evolution of clustering.
Fig.~\ref{fig:zev} shows the results.
The measurements are noisy, because of the low number of galaxies
(comparable to previous studies of optically selected galaxies, 
e.g. Magliocchetti \& Maddox 1999).
A comoving clustering with $r_0\sim3.5$--4.5, that remains
constant up to $z\sim3$--4, is roughly consistent with the measured
 clustering, in
agreement with what is found from the inversion of the angular clustering
of the whole $K<24$ sample.
Our measurements can be compared with the analogous ones
from $I$-selected samples of faint galaxies
in the HDF North (Magliocchetti \& Maddox 1999,
Arnouts et al.  1999) and HDF-S (Arnouts et al.  2002), derived using
photometric redshifts.  Even though a detailed comparison as
a function of redshift cannot be obtained, we notice that
the average comoving correlation lengths of $r_0\sim2$--3 \h1 Mpc
measured for the optically-selected galaxies is
lower than the average one derived from our $K$-band sample.

A particularly interesting redshift range is the one at $z\sim3$, where
measurements of clustering for LBGs samples are available as well.
In the HDF-S, Arnouts et al.  (2002) measure $A=9.6\pm3.0\times 10^{-4}$ 
and $r_0=3.2\pm0.7$ \h1 Mpc for $2.5<z_{phot}<3.5$, 
whereas we find  $A=36\pm8.1\times 10^{-4}$ and $r_0=6.5^{+0.8}_{-0.9}$ \h1 Mpc
in the same redshift range.
Therefore, $K$-selected galaxies appear to be more clustered at $z_{phot}\sim3$
than $I$-selected ones, at the $3\sigma$ level of
significance.
 
An interesting and more accurate comparison can be performed 
with the clustering of LBGs at $z=3$,
using the most up-to-date estimates from GD01. 
To compare with the LBGs we add together all the $K<24$ galaxies in 
$2<z_{phot}<4$, and find $A=17.1\pm4.6\times 10^{-4}$ and $r_0=5.5\pm 0.8$
\h1 Mpc.

\begin{figure}[ht]
\includegraphics[width=9cm]{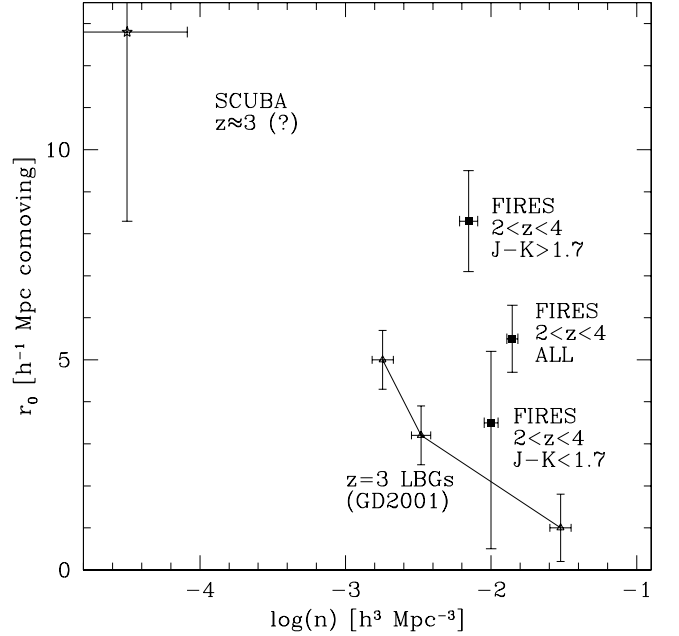}
\caption{Comoving correlation lengths for FIRES galaxies at $2<z_{phot}<4$
(filled squares, see labels for all $2<z_{phot}<4$ galaxies, $J-K>1.7$
and $J-K<1.7$ galaxies),
for LBGs (empty triangles, connected by a line,
from GD01) and for SCUBA galaxies (star, see text for references) 
are plotted as a function of number density. 
}
\label{fig:giava}
\end{figure}

GD01 estimate a clustering amplitude of
$r_0\sim1\pm1$ \h1 Mpc for the fainter HDF sample, with evidence 
that the amplitude depends on galaxy luminosity, and hence galaxy
number density.
Brighter (i.e. rarer)
sources have stronger clustering, reaching $r_0=5.0\pm0.7$
for the so-called "SPEC sample" with ${\cal R}<25$. 
For a fair comparison to LBG samples we have therefore to carefully 
take into account number-density effects. We plot in Fig.~\ref{fig:giava} the correlation lengths as a function of number-density
for $z\sim3$ LBGs and $2<z_{phot}<4$ FIRES galaxies. 
{To derive the number density of our samples we divide the
observed number of galaxies by the volume defined by the area on the sky 
and the FWHM of the photometric redshift distribution}\footnote{
{This corresponds, e.g., to the volume at $2.3<z<3.7$ 
for galaxies with $J-K>1.7$ and $2<z_{phot}<4$.}
We note, however, that systematic effects in the estimate of the sample redshift
distribution would influence both the clustering and number density 
estimates, moving the points in Fig.~\ref{fig:giava} in a direction
basically parallel to the LBG trend, having thus little effect on the
inferred conclusions.}. 
$K$-selected galaxies at $2<z_{phot}<4$, have a 
correlation length of $r_0=5.5\pm0.8$, and are therefore more clustered
than LBGs with similar number density, which have $r_0\sim1.5$--2,
based on the observed scaling (Fig.~\ref{fig:giava}). To derive the significance of this
result, we notice that assuming $r_0\sim1.5$--2 for the FIRES galaxies would
result in an angular clustering amplitude $A\simlt3\times10^{-4}$,
whereas we observe $A=17.1\pm4.6\times10^{-4}$, therefore the effect is
at the $\simgt 3\sigma$ level. 
\begin{figure}[ht]
\epsscale{0.85}
\includegraphics[width=9cm]{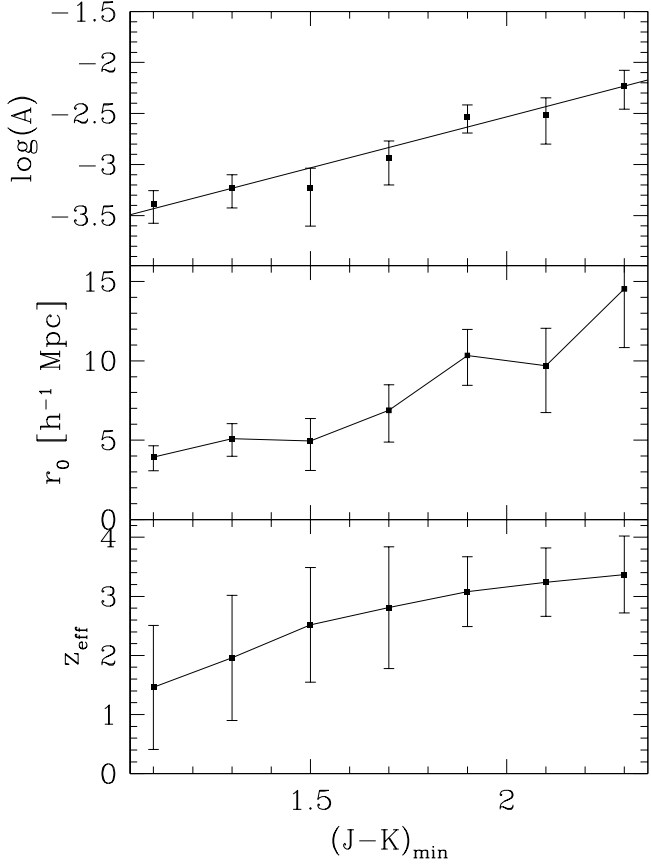}
\caption{
The clustering amplitude (top panel), the inferred comoving correlation length
(center panel, in \h1 Mpc) and the {\em effective} redshift
are given for galaxies redder than the $J-K$ color threshold.
The error bar on the effective redshift
show the standard deviation of the photometric redshift distribution.
The power law fit to the angular clustering (top panel) has 
$log(A)=1.00(J-K)-4.53$
}
\label{fig:JK}
\end{figure}
In conclusion, there is a strong suggestion that $K$-selected
galaxies at $z\sim3$
have either a larger $r_0$ at a fixed number density or a higher number density 
at a fixed $r_0$, with respect to LBGs.

{It is important to stress that this conclusion is robust against 
possible effects due to cosmic variance. In fact, the
fluctuations due to cosmic variance expected in the HDF-S for a clustering
amplitude of $A\simlt3\times10^{-4}$, similar to that of LBGs, are about 10\%
of such a value (Fig. \ref{fig:sn}). 
This cannot produce the much (5 times) larger clustering
amplitude that we have measured.}

\subsection{Clustering of red $J-K$ galaxies at $2<z_{phot}<4$}

\begin{figure}[ht]
\centering 
\includegraphics[width=9cm]{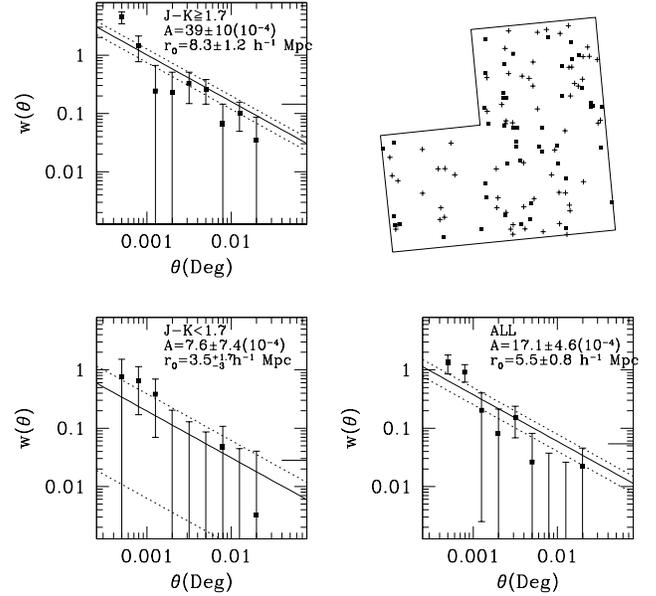}
\caption{The sky distribution of $2<z_{phot}<4$ galaxies (top-right
panel). Objects with $J-K<1.7$ are shown as crosses, while filled
squares are for the ones with $J-K\geq1.7$. The remaining panels show
the measured two-point correlation functions for the different
subsamples at $2<z_{phot}<4$.
}
\label{fig:z3jk}
\end{figure}

An efficient approach to study the clustering of high redshift (i.e. $z>2$)
galaxies is to select galaxies by their $J-K$ color. The effective
redshift of $J-K$ color selected samples grows to $z\sim3$ for the
reddest thresholds.
The clustering amplitude also increases with increasing $J-K$
color threshold (Fig.~\ref{fig:JK}).
The resulting comoving correlation length is also
a strong function of color, again with values of $r_0\sim10$ for 
$J-K>1.7$--2.3, pointing to the interesting result
that a very strongly clustered population of galaxies exists at $z\sim3$.

\subsubsection{Color segregation of clustering in $2<z_{phot}<4$}

We have measured the angular clustering of galaxies restricted to those
with photometric redshift estimate $2<z_{phot}<4$. Dividing the
sample at $J-K=1.7$ we obtain two subsamples of nearly equal size,
containing 49 and 66 galaxies for the red and blue subsamples
respectively.
For the redder subsample with $J-K\geq1.7$ we obtain $r_0=8.3\pm1.2$ 
(Fig.~\ref{fig:z3jk}), consistent with the trend plotted in Fig.~\ref{fig:JK}. 
If different redshift ranges are chosen within $2<z_{phot}<4$, and/or a $J-K$
threshold redder than 1.7 is used, the derived correlation lengths are 
always in the range 8--10 \h1 Mpc, albeit with larger statistical error. 
The sample with $J-K<1.7$  is much less clustered, with
 $r_0=3.5^{+1.7}_{-3.0}$ (Fig.~\ref{fig:z3jk}).
The difference in angular clustering between the $J-K$ red and blue samples
is significant at the $\sim 3\sigma$ level.
The strongly clustered sample with $J-K>1.7$ appears to be the cause of
the significantly larger clustering measured at $2<z_{phot}<4$ for K-selected
versus optically selected samples or LBGs. The clustering of blue
$J-K<1.7$ galaxies at $2<z_{phot}<4$ is in fact consistent to that of LBGs 
with similar number densities (Fig.~\ref{fig:z3jk}).

The key property for identifying this $z\sim3$ strongly clustered population  
appears indeed to be the presence of very red 
$J-K$ colors, irrespective of the properties of the optical continuum
or of the observed near-IR or optical magnitudes.
We note that the median K-band magnitudes of the red and blue
samples are respectively $K=22.7$ and $K=22.8$, a difference that is
clearly not significant. The median $V_{606}$ magnitude of the $J-K$ redder
and bluer subsample does appear instead to be significantly different, with 
$V_{606}=26.4$ and $V_{606}=25.6$ for the red and blue samples
respectively. This difference in $V_{606}$, which is a consequence
of the $K$-selection of the sample, is not the cause of the measured
clustering segregation, as it would bias the bluer-brighter
(and not the redder-fainter) 
subsample toward larger clustering (GD01).
Finally, we tested that no measurable color dependence of clustering can be 
found if galaxies at $2<z_{phot}<4$ are split by their $U-V$, $V-J$ or even
$V-K$ color.

\begin{figure}[ht]
\includegraphics[width=9cm]{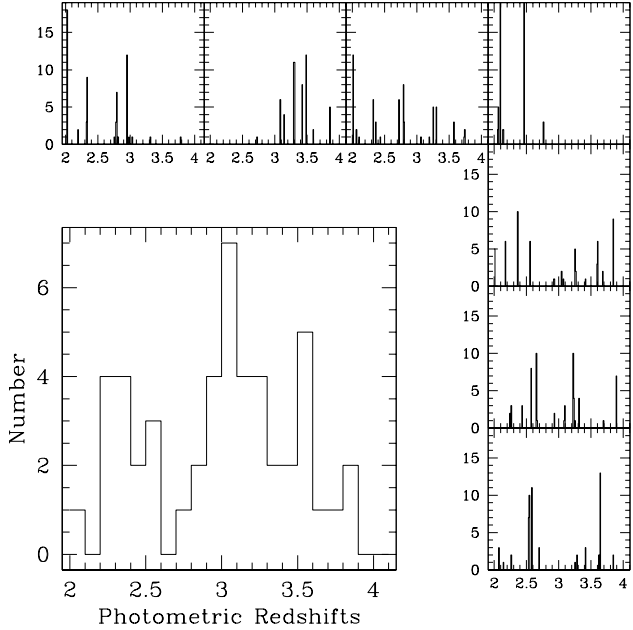}
\caption{The photometric redshift distribution (in 0.1 redshift bins)
of $J-K>1.7$ galaxies at $2<z_{phot}<4$ {is compared to random
(Monte Carlo) realizations of a population at $2<z<4$ with $r_0=8$ \h1
Mpc (small boxes, $\Delta z=0.01$ binning). 
When account is made for photometric redshift
uncertainties, the observed distribution appear consistent with the 
Monte Carlo realizations, i.e. it appear to contain redshift spikes.} 
}
\label{fig:dndz3}
\end{figure}

The large clustering inferred for the $J-K$ reddest 
galaxies at $2<z_{phot}<4$, together with the small area of the HDF-S field, 
consistently imply a highly non uniform redshift distribution 
(see e.g. Broadhurst et al.  1992; Cohen et al 1996; 
Daddi et al.  2001; 2002). To investigate this aspect, 
Monte Carlo simulations were used to produce
random realizations of our survey. A population of galaxies 
with $r_0=8$ \h1 Mpc was extracted with a
flat selection function between $2<z<4$ over a 4 arcmin$^2$ area, 
following the recipes of Daddi et al. 2001. The resulting redshift
distributions (Fig.~\ref{fig:dndz3}) are very spikey.
Fig.~\ref{fig:dndz3} also shows the observed 
photometric redshift distribution of  
galaxies with $J-K\geq1.7$ and $2<z_{phot}<4$, which suggests the
presence of three main galaxy concentrations at $z_{phot}\sim2.4$, 3 and 3.5,
in qualitative agreement with our Monte Carlo simulations.
While systematic errors in the photometric redshift determination may
still result in an incorrect determination of the real redshift of the
spikes, there is evidence that these features reflect
real redshift segregations. In fact, we
tested that restricting the clustering measurements to 
redshift ranges covering such features imply an enhanced angular clustering. 
This effect is expected only if the galaxies in the photometric
redshift spikes are spatially associated.

As a further confirmation, van Dokkum et al.  (2003) 
present in a companion paper spectroscopic
confirmation of 5 galaxies with $J-K>2.3$ at $2<z<4$ (selected
from the MS1054-03 field), with 3 of them belonging to a single redshift
spike, in qualitative agreement with the estimates of large clustering 
presented here.
We also note that Saracco et al.  (2001) using 
IR data on HDF-S and "Chandra Deep Field South" found that
$J-K>2.3$ galaxies are unevenly distributed, with their surface
density varying by over a factor of two over the two fields, 
qualitatively consistent with the results presented here.


In Fig.~\ref{fig:z3jk} (upper-left panel) an excess of very close pairs 
(with separations $\sim2$ arcsec) 
of $J-K$ red galaxies is apparent in the measured two point correlation
function.
A similar excess at short separations was noted by Ouchi et al. (2001) in their 
sample of $z\sim4$ LBGs.
This feature is potentially interesting: 2 arcsec at $z\sim3$ corresponds
to $\sim50$ kpc and suggest a merging destiny for these galaxies (if they are
indeed at identical redshifts).
We find that the excess mainly arises because of 2
triplets of very close separation galaxies in our sample, 
yielding a large number of
close pairs: 5 of these have photometric redshift estimates placing
them in the $z_{phot}=3.5$ spike of galaxies and $J-K>2.1$ colors (i.e. at the extreme
range). The feature disappears if the clustering is
measured excluding these "triplets" galaxies. 
At the same time, such an excess at a very small scale may
suggest a correlation function slope steeper than the one assumed 
($\gamma>1.8$), which would be
consistent with measurements for local early-type galaxies (Guzzo et al.
1997). {Our measurements cannot significantly
constrain the slope of the correlation functions, because of
the relatively small range of probed angular scales.
Previous measurements for LBGs suggest possible values of $\gamma=1.5$--2.1
(Giavalisco et al. 1998, GD01, Porciani \& Giavalisco 2002) that would
all be basically consistent with our data.
We notice, however, that for the case of $J-K>1.7$ galaxies in
$2<z_{phot}<4$ (with estimated $r_0=8.3$ \h1 Mpc for $\gamma=1.8$), changing
$\gamma$ to 1.5(2.1) increase(decrease) the inferred $r_0$ by only
7.5\%(8.5\%).}

\subsubsection{$J-K$ red galaxies at $2<z_{phot}<4$ and LBGs}
\label{sec:LBG}

\begin{figure}[ht]
\includegraphics[width=9cm]{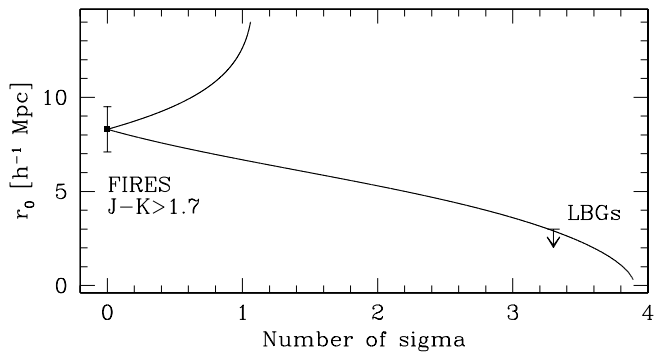}
\caption{Significance levels for the clustering of faint $K$-selected 
galaxies with $J-K>1.7$, accounting for both the measurement error and the 
cosmic variance fluctuations. The correlation length is greaten than 5.3(3.6)
\h1 Mpc at the 2(3) sigma level. A limit on the correlation length of LBGs
with similar number density is plotted for comparison.
}
\label{fig:error}
\end{figure}

It is relevant at this point to compare in detail 
the number density and clustering
of this strongly clustered $z\sim3$, $J-K>1.7$ population to 
the LBGs (GD01, see also Adelberger et al.  1998, Giavalisco et al. 1998, Porciani \& Giavalisco 2002). 
LBGs with a comoving density similar to this 
population are expected to have $r_0\sim2$--2.5 (Fig.~\ref{fig:giava}), 
that would produce an angular clustering of $A\simlt5\times 10^{-4}$,
assuming the observed photometric redshift distribution of our red $z\sim3$ 
population. 
In contrast, for the red $K$-selected galaxies at $z\sim3$, 
$A=38\pm10\times 10^{-4}$
is measured. 
Hence the red galaxies are more clustered than LBGs with the same
number densities at the $3.3\sigma$  level.
{The results appear again robust with respect to
cosmic variance in the clustering. To quantitatively address the point we 
analyze in Fig. \ref{fig:error} the constraint on the {\em average}
correlation length of
$K$-selected galaxies with $J-K>1.7$ colors, allowing for both the measurement
error and the cosmic variance fluctuations (added in quadrature).}

The selection of FIRES galaxies is  based solely on the
$K$-band apparent magnitude. 
The standard $z\sim3$ LBG selection is based instead 
on the presence of an $U$-band dropout and blue
optical colors, which in principle is biased against galaxies with 
low star-formation rates or those that are highly obscured by dust.
It is important, therefore, to compare the properties of  
the samples selected by the two criteria.

We adopt the HDF-LBG definition by
GD01,\vspace{.4truecm}\\
$(U_{300}-B_{450})\geq 1.0+(B_{450}-V_{606}), $\\
$(U_{300}-B_{450})\geq 1.6,\ \ (B_{450}-V_{606})\leq 1.2$
\vspace{.4truecm}\\
(AB magnitudes). Restricting ourselves to the redshift range where the LBG
selection is effective, i.e. $2.0<z_{phot}<3.3$, we find 
that about 30\% (10/35) of the $J-K>1.7$ galaxies 
satisfy these criteria. The remaining galaxies are still
consistent with being LBGs, all having quite blue 
$0<(B_{450}-V_{606})_{AB}<1$ colors (within $2\sigma$), but are
very faint in the optical and have lower limits to the
$U_{300}-B_{450}$ colors that do not allow them to be
individually classified as LBGs even in the very deep HDF-S data. 
However, we find that, by
stacking together those that do not individually qualify as a LBG,
we obtain average colors fully consistent with those of a $U$-dropout galaxy,
i.e. an LBG. 

Even if the $J-K$ red galaxies at $2<z_{phot}<4$ and the LBGs may
not be distinguishable from the rest-frame UV colors, the two criteria
define populations with quite different properties
for at least two main reasons: significantly different
clustering properties and significantly different rest-frame optical
colors (i.e. the observed $J-K$). 
The overlap between our sample of $z\sim3$ red galaxies 
and the LBG samples selected in the HDF-S by GD01 to $V_{606}<27$
is in fact small.

The color segregation of clustering between $J-K$ red galaxies on
one side, and bluer $K$-selected galaxies or LBGs on the other side,
appears as a much stronger effect than the luminosity segregation
among LBGs discussed by GD01. This is similar to what
observed for local galaxies, where color (i.e. morphological)
segregation of clustering is strong and well established at least
 since Davis \& Geller
(1976), while a convincing measure of the mild luminosity dependence of
clustering at $z=0$ has been achieved only in recent years (Norberg et
al.  2002).

\subsubsection{$J-K$ red galaxies at $2<z_{phot}<4$ and SCUBA sources} 

Sub-mm bright galaxies detected by SCUBA are another interesting
classes of sources of which a significant fraction is expected to be at
$z>2$ (e.g. Smail et al. 2002a and reference therein).  Scott et
al. (2002) and Webb et al. (2002) recently report a $\approx2\sigma$
detection of angular clustering, and infer a correlation length of
$r_0=12.8\pm4.5$ \h1 Mpc.  Such high level of clustering is consistent
with what we find for the red  $J-K$ galaxies at $2<z_{phot}<4$.

However, the number density of SCUBA sources is estimated in the range
$10^{-4}$--$10^{-5}$ h$^3$ Mpc$^{-3}$ (Scott et al. 2002; Smail et al
2002b), about 2 order of magnitudes lower than that of the red
$J-K$ galaxies at $2<z_{phot}<4$ (Fig.~\ref{fig:giava}). 
Given the large difference in these number
densities, it is possible that these SCUBA sources are a subset of the
red $J-K$ galaxies.
\begin{figure}[ht]
\includegraphics[width=9cm]{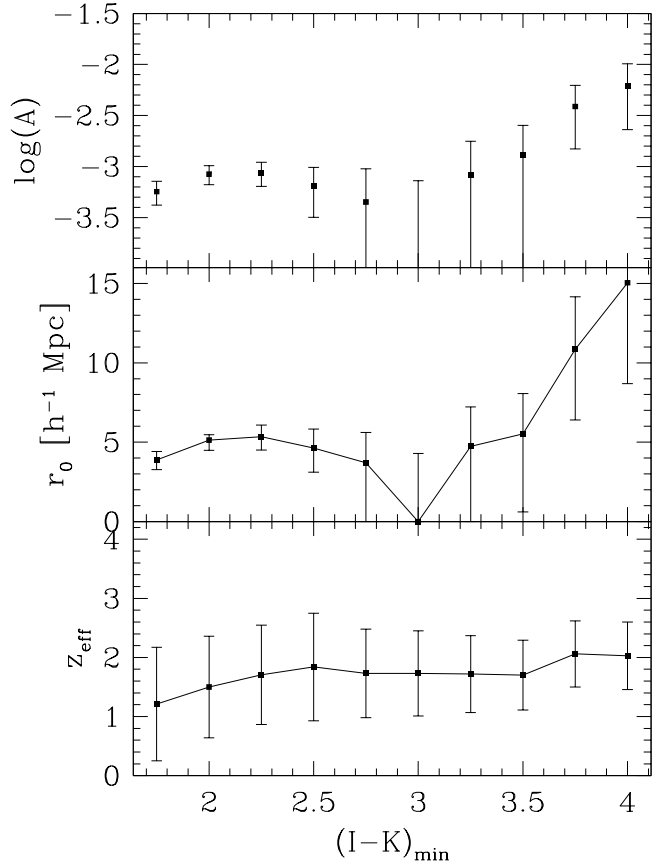}
\caption{ 
The same of Fig.~\ref{fig:JK} but for the $I-K$ color threshold. 
}
\label{fig:IK}
\end{figure}

Studies of SCUBA galaxies so far have mainly concentrated on
their $R-K$ and/or $I-K$ colors (i.e. checking if they
qualify as EROs), often yielding ERO--like colors.  It would be
very interesting to determine the $J-K$ colors of a statistical sample
of SCUBA sources to
test whether the SCUBA galaxies are indeed a subset of the new class
of $J-K$ red objects highlighted by our deep $K$-band FIRES survey.

\subsection{Clustering of faint EROs at $0.8<z_{phot}<2$}

For completeness, we present in this paragraph the clustering of $K$-selected
galaxies as a function of the $I-K$ color. Galaxies with red $I-K$
colors are classified as extremely red objects (EROs), and allow one
to address redshift ranges at $1\simlt z \simlt 2$ (Cimatti et al. 2002a),
in agreement with our photometric redshift estimates.
Daddi et al. (2000b) found a monotonic power-law increase of $log(A)$ as
a function of $R-K$ color for $K<18.8$ galaxies.  For the FIRES
galaxies that are on average 5 magnitudes fainter we still detect a
general correlation between clustering and $I-K$ color.
(Fig.~\ref{fig:IK}). A difference is that for the FIRES galaxies the
clustering amplitude shows a plateau for $2.5<I-K<3.5$ and increases
again toward redder $I-K$ colors.  Using the photometric redshifts
we find that the angular amplitude implies a correlation length
consistent with $r_0\sim4$--5 \h1 Mpc for the galaxies with $I-K$
selection threshold lower than 3.5, applied at {\em effective}
redshifts $1<\overline{z}<2$ (see also Fig.~\ref{fig:IK}).

\begin{figure}[ht]
\includegraphics[width=9cm]{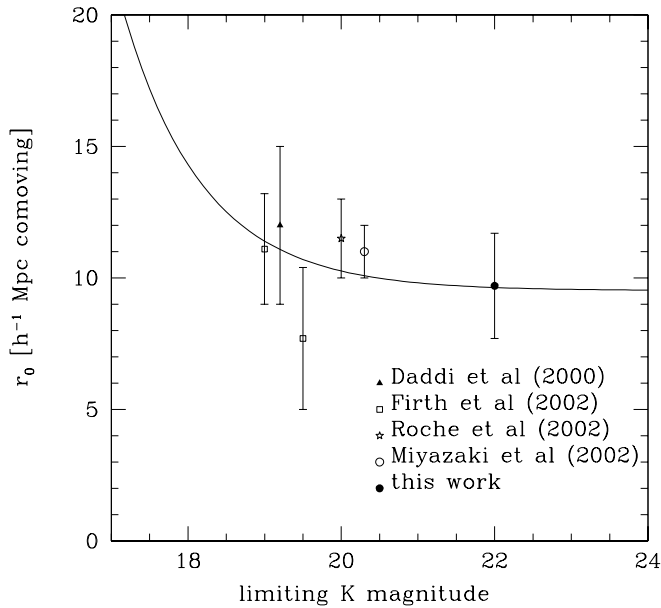}
\caption{The comoving correlation length inferred for EROs
($z\approx1$), as a
function of the sample limiting magnitude. 
The fitted line is based on the scaling $b/b^*=0.85+0.15*L/L^*$ (Norberg
et al. 2002). 
}
\label{fig:nn}
\end{figure}

Galaxies with $I-K>3.5$ show 
a stronger amplitude, and are consistent with $r_0\simgt8$ \h1 Mpc.
The $I-K>3.5$ selection criterion is close to the ERO selection criterion 
$R-K>5$. 
Even though our measurements extend in principle up to 5 magnitudes deeper,
very similar values are found for the clustering of $z\sim1$ EROs
(Daddi et al.  2001, 2002; Firth et al.  2002, Roche et al.  2002,
{Miyazaki et al. 2002}). 
However, we note that the clustering signals here predominantly arise from 
red galaxies with $1<z_{phot}<2$ that have typically $K<22$ 
and are therefore only
$\sim2$ magnitudes fainter than the EROs analyzed by Roche et al.  (2002)
and Miyazaki et al. (2002).

The K20 survey (Cimatti et al. 2002a, Daddi et al. 2002) established from 
deep VLT spectroscopy that $K\sim19$--20 EROs come in two different
flavors: early-type galaxies and reddened starbursts. About 12(3) of the
FIRES galaxies with $I-K>3.5$ have $K<20(19)$ and would have been inserted
in the K20 sample, i.e. only 15\%(4\%).
Recent studies of the faint $K>20$
ERO population suggest a non-negligible fraction of dusty starbursts
(Wehner et al. 2002, Smail et al. 2002b).
Our faint $I-K>3.5$ EROs at $z_{phot}<2$ typically 
have $1.6<J-K<2.0$, which may favor these to be predominantly 
early-type galaxies (Pozzetti \& Mannucci 2000). 
No galaxies with $I-K>3.5$ and $z_{phot}<2$ are found at $K>22$. This is
probably not an effect of cosmic variance, as Yamada et al. (2001) notice 
a similar behavior at $K>22$ in the HDF-N and 53W002 field. 
A low density of $K>22$ EROs would be expected assuming
a "bell shaped" early-type galaxy luminosity function, 
a feature observed up to $z=1$ (Pozzetti et al. 2002, cfr. also Totani 
et al.  2001 and Smail et al. 2002b on this point).
Again,  this behavior is expected  for faint early galaxies, 
but seems unlikely for starbursts.
In addition, Daddi et al.  (2002) present evidence that the ERO clustering 
at $z<2$ is mainly due to the early-type galaxies. The dusty EROs are found
to be very weakly clustered, but their diluting effect on the overall 
clustering is weakened by a non-negligible cross-correlation between
the two galaxy types {(see however Miyazaki et al. 2002 for a somewhat
different result)}.

{Even if the uncertainties here are much larger 
than for $z>2$ red galaxies discussed in the previous sections
(mainly because the clustering amplitudes for EROs are measured 
at the $\approx2.5\sigma$ level of significance only)}
we find evidence, in conclusion, for faint $L\simlt0.1L^*$\footnote{assuming $K(L^*,z=1)\sim19$ 
from Daddi et al. 2000a} early-type galaxy clustering that is still consistent
with a large $r_0$ of 10 \h1 Mpc at $1\simlt z\simlt2$. 
This high level of clustering of faint EROs suggests
a weak luminosity dependence of early-type galaxy clustering at $z\sim1$. 
In the local
universe it has been recently found that the bias of galaxies (and in particular
also of early-type galaxies) follows the relation
$b/b^*=0.85+0.15\times (L/L^*)$ (Norberg et al. 2002). We plot in Fig.~\ref{fig:nn} the EROs clustering measurements taken from the literature
together with the best-fitting Norberg-like law. 
For the FIRES EROs we report the measurements obtained for galaxies with
$I-K>3.5$ and $0.8<z_{phot}<2$, i.e. $r_0=9.7\pm2$ \h1 Mpc.
Only the Daddi et al.  (2001) data explicitly take into account 
cosmic variance of clustering, while it is actually the highest precision
measurement ($S/N\sim8$, vs $S/N\sim2$--3 for most of the other measurements).
The observed trend is in agreement with the local scaling
behavior of the clustering with luminosity. Our new data at faint
luminosity is consistent with the large overall normalization of the
relation, with $r_0(L=L^*, z\sim1)\sim 11$ \h1 Mpc, larger than
local values. This again favors the hierarchical clustering scenario for
the evolution of early-type galaxies clustering (see Daddi et al. 2001
for a more detailed discussion of this point).
In principle, because of clustering evolution, the local relation for the 
luminosity dependence of the bias may be
expected to change with redshift. In particular, at $1<z<2$, a
steeper dependence on luminosity should be expected at the bright end
($L>L^*$), both for "galaxy conservation" and "galaxy merging"
scenarios. This cannot be ruled out, of course, on the basis of the
present data: the $K=19$ measurements were in fact shown to be
lower limits, because of the diluting effect of the dusty
EROs on clustering (Daddi et al.  2002). 
Anyway, it seems probable that a quite large $r_0\simgt 15$ \h1 Mpc may be
expected for very bright $K\sim17$--18 early-type galaxies at $z\sim1$.
It is natural to identify these brightest EROs with the short lived
powerful radio galaxies that have $r_0$ consistent with high values
(Overzier et al. 2002).

\section{Discussion and modeling}
\label{sec:models}

\subsection{An empirical model for the clustering at faint $K$ magnitudes}

We will first discuss the evolution of the clustering of galaxies
selected in the $K$-band. With a number of simple assumptions we will
first model the relatively flat dependence of the angular clustering
amplitude on the $K$ magnitude down to the faint $K=24$ magnitude (see
Fig.~\ref{fig:roche}).

Galaxies with $K$-magnitudes in the range $15<K<18$,  i.e. down to
the limit where the slope of the amplitude versus K-band magnitude
changes, have a median redshift that scales with $K$-magnitudes and is
in the range of 0.1 to 0.4 (Songaila et al
1994, Cowie et al. 1996). The number counts in this range can be
directly predicted on the basis of the local luminosity function
(Kochanek et al. 2001). In this regime the dependence of the
clustering amplitude on limiting magnitude can be described using a
simple scaling relation as a function of depth reached (Peebles 1980).
This is due to the fact that,
within this limited redshift range, clustering evolution
is very mild (Baugh et al. 1996).  At fainter magnitudes the tail of
high redshift galaxies becomes significant (Fig.~\ref{fig:distri})
allowing studies of the redshift evolution of clustering.

Assuming a comoving clustering of $r_0\sim4$ \h1 Mpc at all redshifts
(consistent with Fig.~\ref{fig:zev}) and using the observed $N(z)$, we
can reproduce correctly the inferred trend of amplitude versus $K$
magnitude of Fig.~\ref{fig:roche}.  
However, faint K-band selected samples are a mix of at least two
different populations with different clustering evolution properties.  In
addition to the strongly clustered red galaxies it contains a
significant fraction of poorly clustered blue 
galaxies in common with the optically-selected ones. 
McCracken et al.  (2001) find that the angular clustering amplitude
as a function of the $I$-band magnitude 
($22<I_{AB}<25$) can be reproduced by a strongly evolving population with
$r_0(z=0)=4.3$ \h1 Mpc and $\epsilon=1$. Here they used the standard 
"empirical" parameterization (e.g. Peebles 1980)
for the evolution of the 2-point correlation 
function $\xi(r,z)=\xi(r,0)(1+z)^{\gamma-3+\epsilon}$, where the free
parameter $\epsilon$ determines the clustering evolution.
In this scenario the comoving correlation length 
at $z\sim1$ has already dropped by 60\% to a value of $r_0\sim1.8$ \h1 Mpc,
consistent with measurements at this redshift by the CFRS (LeF\`evre et
al.  1996), and decreases even further at higher $z$.
Adopting the McCracken et al.  evolution for the clustering and
using the observed photometric redshift distribution in $20\leq K\leq24$,
Fig.~\ref{fig:roche} (dotted line) predicts clustering at
faint $K$ magnitudes which is nearly an order of magnitude lower than 
observed in the FIRES survey. 
The McCracken et al.  parameterization 
of clustering for faint $I$-band selected galaxies would also be
inconsistent with the evolutionary trend inferred for $r_0$ as a
function of $z$ (Fig.~\ref{fig:zev}).

$K$-band selected samples at high-$z$ must therefore contain 
a more clustered population, 
to boost the $K$-selected correlations, in addition to the galaxies
in common with optically selected samples.
It is natural to identify massive early-type galaxies with this population.
Due to an intrinsically old and passive stellar population
leading to large K-corrections, these galaxies have red 
optical to near-IR colors at least to $z\sim2$.
Early type galaxies should thus be more prominent in near-IR selected samples,
while they could escape optically selected samples of comparable
depths.

Moreover, the clustering of red early-type galaxies at
$z\sim1$ has been recently measured (Daddi et al.  2000b, 2001, 2002, 
Firth et al.   2002, Roche et al.  2002) 
to be very strong with $r_0\simgt10$ \h1 Mpc, a
value comparable to or even greater than that in the local universe, 
consistently with this paper's findings.

In the light of these recent developments, 
we calculate predictions for the $K$-magnitude dependence of clustering
based on a two component model containing: (1) a population of
early-type galaxies with a redshift distribution derived assuming a
negligible number density evolution from $z=0$, following the
Daddi et al.  (2000a) model and with a constant comoving clustering 
of $r_0=10$ \h1 Mpc, involving typically 15\% of the galaxies at
$K=20$--24; and (2) the remaining galaxies with the rapidly decreasing
clustering derived by McCracken et al.  (2001), having 
$r_0(z=0)=4.3$ \h1 Mpc and $\epsilon=1$. The cross correlation between 
the two samples is assumed to have a correlation length equal 
to the geometrical mean of the ones of the two distinct populations 
(as expected in the case of two differently biased realizations of the same 
underlying matter distribution).
We show in Fig.~\ref{fig:roche} that this two component model reproduces
the correlations very well, suggesting that a galaxy population with
large ($r_0\sim10$ \h1 Mpc) clustering extending to high redshift ($z\simgt3$)
can explain the observed flattening.
We recall that our measurements of color selected subsamples of 
$K<24$ galaxies confirm 
the presence of such strongly clustered populations at $1\simlt z\simlt 4$.

\subsection{On the nature of the strongly clustered population at $z\sim3$ with
$J-K>1.7$}

The main result of this study is the detection of a
population with $r_0\sim8$ \h1 Mpc at $z\sim3$. 
We recall that there is much supporting evidence for the
existence of such a highly clustered population:
\\
$\bullet$ The large clustering amplitude measured for all galaxies 
to $K=24$ requires the presence of galaxies with strong clustering at
$z>2$.\\
$\bullet$ The clustering of $2<z_{phot}<4$ $K$-selected 
galaxies is high, and larger than 
that of optically selected HDF galaxies and of LBGs over similar redshift
ranges and with similar number densities.\\
$\bullet$ We detect a well defined trend of increasing angular clustering with
an increasing minimum $J-K$ color threshold for the whole $K<24$ sample, 
measurable up to the threshold of $J-K>2.3$. 
With the use of photometric redshifts we find that such a trend implies 
a growth of the spatial clustering with 
$r_0\simgt8$ \h1 Mpc for $J-K>1.7$, again holding at an effective redshift
$z\sim3$\\
$\bullet$ Preselecting galaxies with photometric redshifts $2<z_{phot}<4$, 
we detect significant color dependence of clustering at those redshifts. 
If the sample is split at $J-K=1.7$,  $r_0=8.3\pm1.2$ \h1 Mpc
for the $J-K>1.7$ sample at $2<z_{phot}<4$\\
$\bullet$ We find apparently significant 
spiky structure at $2<z<4$ in the photometric redshift 
distribution, as required by the large inferred correlation length.\\
$\bullet$ There is evidence of strong field to field variation of faint $J-K$
red galaxies, between HDF-S, HDF-N and CDFS (Labb\'e et al. 2003, Saracco et
al. 2001, this work)\\
$\bullet$ There is preliminary confirmation of strong redshift space 
clustering of red $J-K$ galaxies at $z>2$ (van Dokkum et al. 2003)

The question that now arises is which is the main origin of the observed
$J-K>1.7$ colors of this strongly clustered population. 
The redder $J-K$ galaxies may be older and/or 
more massive, may be heavily reddened by dust or may have prominent lines
(e.g. $H_\alpha$) falling in the $K$ band.  
Indeed, van Dokkum et al. (2003)
detect emission lines in 4/5 of the bright $J-K>2.3$ spectroscopically
confirmed objects at $z>2$, and in particular AGN lines in 2/5. 
Their contribution to the broad band colors
is however estimated to be small, $\simlt10$\%.
Furthermore, the objects with red $J-K>1.7$ colors at $2<z_{phot}<4$ have
generally a relatively flat and blue spectral energy distribution
blueward of the $J$-band (with median {$(B-I)_{AB}=1.01$ fully 
consistent with
the (B-I) colors of ordinary LBGs}), suggesting the
presence of a break between the $J$ and $K$ bands and disfavoring 
strong dust reddening in most cases.
We notice that at $2<z<4$ the J- and K- bands 
sample around the rest frame 3000 {\rm \AA}\ and 5500 \AA\ regions
respectively, thus targeting a region around the 4000 \AA\ break.
This suggest that the most likely interpretation of the peculiarity of
this $J-K$ red population is an age effect.  This agrees with a detailed 
modeling of the SEDs of the reddest ($J-K>2.3$) galaxies with $z_{phot}>2$, 
that is presented in a companion paper (Franx et al. 2003).

Recently, some modeling of the properties of galaxies at $z\approx3$ have also been presented for LBG samples having both near-IR imaging and spectral
observations available (Shapley et al.  2001, Papovich et al.  2001).
Shapley et al. describe both $J$- and $K$-band
observations, thus allowing for comparison with our work, and estimate
the best fitting properties of a sample of spectroscopically confirmed 
LBGs at $2<z<3.5$, 
including the dust reddening, star-formation rate, stellar mass and
age (intended as the time elapsed since the onset of a continous 
star-formation). 
We extract a subsample of 40
LBGs from the work of Shapley et al.  (2001) having all the necessary
information available, including the $J-K$ color and spectroscopic redshift.
We find upon dividing the Shapley et al. galaxies at the
$J-K=1.7$ color, that the reddest galaxies turn out to have a larger median
mass ($3\times10^{10}M_\odot$ vs $10^{10}M_\odot$), larger median
age (572 Myr vs 286 Myr) but similar median
star-formation rate ($\sim45M_\odot/yr$) and extinction properties
(E(B-V)$\sim0.17$). This is in agreement with our qualitative
considerations presented above.
A Kolmogorov-Smirnov test shows that the difference in the 
mass distributions in the two samples is significant at the $>99$\%
confidence level, 
while for the age the significance is at the 85\% level. It should
also be noted, however, that the typical error on the $J-K$ color is
quite large ($\simgt 0.4$ magnitudes), resulting in some mix in the
samples, so that the intrinsic trends and differences could be even stronger
than found.

While doing this comparison, we have to keep in mind that
these Shapley et al.  LBGs are brighter than ours (with a median $K\sim21.4$ 
versus $K\sim22.7$ for our $2<z_{phot}<4$ galaxies), and that the
Shapley et al. sample was optically selected, and subsequently observed in the
near-IR.
The differences in the mass and {\it age} for $J-K$ red and blue
subsamples of LBGs are in fact a consequence of strong correlations of
the fitted mass with the apparent $K$ magnitude (i.e., more massive LBGs
are brighter in
$K$) and of the fitted star-formation rate with the 
optical magnitudes (LBGs with larger star-formation rates are
brighter in, e.g., {\em R}). 
Rather interestingly, the Shapley et al.  (2001)
modeling ultimately suggest that the $K$-selected $J-K$ red
galaxies are more massive and relatively older than LBG samples
selected at comparable depth. This imply a low star-formation rate per
unit mass for this population, that appears therefore to be relatively 
evolved and, thus, presumably
formed at the highest density peaks in the matter distribution at
significantly earlier epochs.
Retrospectively, if we have a way to isolate the oldest and most massive
galaxies in a sample (at any redshift), then it is natural to expect a larger
clustering for this population, with respect to the other younger and
less massive galaxies.  This gives a rather consistent way of
interpreting the significantly larger clustering of $J-K>1.7$
galaxies at $2<z_{phot}<4$, with respect to LBGs and bluer $J-K<1.7$ objects.

Our findings suggest therefore that a color-density relation, similar to that
observed in the local universe (i.e. driven by the star-formation rate per unit mass, Dressler 1980), was in place since early epochs (at least at $z\sim3$)
and is therefore not simply a product of gravitational growth of
clustering at lower redshift.

\subsection{Theoretical models and the existence of a strongly
clustered population at $z\sim3$}

For the clustering evolution of early-type galaxies 
the semi-analytical hierarchical model by Kauffmann et al.  (1999)
predict a large and nearly constant comoving clustering up to $z\sim3$, 
matching our observational results well, if indeed this strongly
clustered population at $z\sim3$ is evolving into early-type
galaxies.
The same hierarchical models, however, 
require a rapid decline in their number density to
$z\sim3$, at least for the most massive systems. We have explored this
point by analyzing the population of galaxies included in the GIF
simulations\footnote{http://www.mpa-garching.mpg.de/GIF/}
based on the  Kauffmann et al.  (1999) models. The corresponding
simulated catalog at $z=2.97$ is limited to objects with
$M>10^{10}M_\odot$, and we find that populations with large
$r_0\simgt 8$ \h1 Mpc can be selected with different criteria (e.g. the most
massive, or the most star-forming), but these populations are typically
$\simgt10$-100 times less abundant in number density
than the population we have identified. 
Scaling from the mass estimates of Shapley et al.  (2001), and
consistent with our own estimates (Rudnick et al.  in preparation), 
it seems likely that roughly 30--50\% of our $2<z_{phot}<4$, $J-K>1.7$ galaxies 
have $M>10^{10}M_\odot$, suggesting that a discrepancy with the 
Kauffmann et al.  (1999) modeling exists.

In the hierarchical framework a joint analysis of
the clustering and number density of a population of galaxies can be
used to constrain their halo occupation function (e.g. Wechsler et al.
2001).
From Mo \& White (2002), a comoving correlation length of 
$r_0\geq 8$ \h1 Mpc at $z=3$ is expected for dark matter halos with 
$M\simgt 10^{13}M_\odot$ in the $\Lambda$CDM models. Such halos have 
a comoving density 
$\sim10^{-4}$ h$^3$ Mpc$^{-3}$, a factor of several tens 
lower than the number density we estimate for the red $J-K>1.7$ 
galaxies at  $z\sim3$ (Fig.~\ref{fig:giava}). 
Even allowing for a somewhat lower clustering reflecting the
observational uncertainties we must conclude that if our estimates are
correct they can be reconciled with the hierarchical clustering scenario
only if large occupation numbers are characteristic of this population.
This is not unexpected in numerical simulations: for example White et
al.  (2001) estimate the existence of about 10 sub-halos for each DM halo
already at $M\sim10^{12}M_\odot$, with the number of sub-halos increasing
almost linearly with halo mass. However, it is expected from
theory that many of these sub-halos would not produce a visible galaxy
(e.g. Primack et al.  2002).

{On the other hand, the existence of numerous sub-halos could
perturb the two-point correlation function at small scales. In Sect. 
3.3.1 we actually do find hints for a small-scale excess in the
correlation functions that could be due to such an effect. However, a
detailed modeling of these aspects within the "halo model" in the
$\Lambda$CDM scenario is beyond the scope of this paper.}

\subsection{Detecting the progenitors of early-type galaxies ?}

In a recent paper, Moustakas \& Somerville (2002) present a detailed
study of the clustering and number density properties of local giant 
ellipticals, EROs and optically selected LBGs 
trying to establish a possible evolutionary
link between these populations, in the framework of hierarchical CDM
models. 
Interestingly they estimate that both 
local giant ellipticals and EROs are hosted typically by
$M\sim10^{13}M_\odot$ halos and are characterized by non-unity occupation
numbers, while they constrain the characteristic halo mass of
(standard, optically selected) LBGs to be
$M\sim10^{11}M_\odot$ or lower, disfavoring the possibility that 
they are the direct progenitors of the former
classes. This conclusion was also reached by Daddi et al.  (2001) on an
empirical basis.

Moustakas \& Somerville (2002) also provide some predictions for the
properties of the progenitors of present-day ellipticals and $z\sim1$
EROs.
These progenitors are expected at $z\sim3$ to be hosted by
$M\simgt10^{13}M_\odot$ halos (similar to their lower $z$ descendants)
and therefore to show large clustering with 
$r_0$ in the range 7--15 \h1 Mpc, and to be characterized by large halo
occupation numbers. These properties are in remarkable agreement for what
is inferred for $K$-selected $J-K>1.7$ galaxies at $2<z_{phot}<4$.

{The large $R-K$ colors and spectral properties of $z\sim1$ early-type
galaxies (Cimatti et al.  2002a), as well as the
fundamental plane studies of ellipticals up to $z<0.8$ (van Dokkum et al.
1998), suggest that  a substantial fraction of the stars ending up in 
local ellipticals and EROs were already in place at $z\simgt 3$. 
This is consistent with, and requires, the existence of relatively old and 
massive galaxies at $z>2$, that we may have detected.}

It is suggestive to interpret both the empirical and the theoretical
suggestions  concluding
that we may have located the progenitors of local massive early-type galaxies 
and of $z\sim1$ EROs in the subsample of very red, $K$-selected,
$J-K>1.7$ galaxies at $z\sim3$.
{As the inferred halo occupation number as well as
the number density of our faint red $z>2$ galaxies are larger than that 
of massive early-type galaxies at $z=0$ or of $z\sim1$ bright EROs, 
merging could be required to reduce both,}
and the total mass of each finally formed galaxy would correspondingly grow.

\section{Summary and conclusions}
\label{sec:concl}

The clustering properties of $K\leq24$ galaxies in the HDF-S field
 have been analyzed.  The main results can be summarized as follow:
\\
$\bullet$ We have produced a first assessment of the clustering of galaxies
as a function of $K$ magnitude up to $K=24$. Whereas the clustering
amplitude flattens, as already known, down to magnitude $K\sim19$, 
it then remains surprisingly high with only a slight further decline
out to $K=24$.
\\
$\bullet$ Modeling of the clustering of $K$-selected galaxies at $20\leq
K\leq24$ requires a strongly clustered sub-population and is
consistent with a picture in which early-type galaxies are strongly
clustered, with $r_0\approx10$ \h1 Mpc \h1 Mpc, all the way to $z\sim3$. 
\\
$\bullet$ We have analyzed the clustering of galaxies in photometric
redshift bins, and detected strong clustering at $2<z_{phot}<4$.
The clustering of the $K$ band selected sample is stronger than
that of LBG's of similar number density.
\\
$\bullet$ At redshifts higher than 2, the clustering amplitude depends
on the $J-K$ colors
of the galaxies. Redder galaxies have stronger clustering.
Galaxies with $J-K > 1.7$ and $2 < z_{phot} < 4$ have $r_0\sim8$ \h1 Mpc.
These galaxies likely have high ages and high mass-to-light ratios
(see also Franx et al. 2003).
\\
$\bullet$ The color dependence of the clustering suggests that a
color-density relation, qualitatively similar to that observed locally,
was already in place at those early epochs.
\\
$\bullet$ A redshift distribution with prominent spikes is predicted for
the $K$-selected population of HDF-S galaxies with $2<z_{phot}<4$,
particularly for the redder galaxies with $J-K>1.7$, in agreement with our
photometric redshift analysis.
\\
$\bullet$ In a CDM framework, these $J-K$ red galaxies at $2<z<4$ would
be hosted by $M\simgt 10^{13}M_\odot$ halos, with large occupation
numbers (i.e. within sub-halos).
Semi-analytical models appear to severely underestimate the number 
density of strongly clustered, $r_0\simgt8$ \h1 Mpc, 
galaxies at $z\sim3$, that are as numerous as faint LBGs.
\\
$\bullet$ We have discussed the properties of this newly discovered
population of strongly clustered $z\sim3$ galaxies, including number
densities and clustering, with the plausible conclusion that a direct
evolutionary trend exists between these $J-K$ red $z\sim3$ galaxies 
on one side and EROs at $z\sim1.5$ and local massive early-type galaxies
on the other side.
\bigskip

{
Over the last several years the popular scenario in which bulge-dominated
galaxies form at relatively low redshift by the merging of full-sized spirals
has encountered increasing difficulties in accounting for the growing body
of observational evidence (see e.g., Renzini \& Cimatti 1999; Peebles
2002).  In particular, in semi-analytical simulations this scenario
fails to produce enough red galaxies at $z\simgt 1$ (Daddi et
al. 2000a; Smith et al. 2002; Cimatti et al. 2002a) as well as enough
luminous galaxies at $z\simgt 1.5-2$ (Cimatti et al. 2002b; van Dokkum
et al. 2003). In addition, semi-analytical models fail completely
to reproduce the small age difference between early-type galaxies in clusters 
and the field out to $z = 0.6$ (van Dokkum et al. 2001).
Our results may suggest an alternative scenario for the
formation of early-type galaxies, in which they would result from the
rapid coalescence (multiple merging) of the red galaxies at $2<z<4$,
assuming these red galaxies are indeed grouped within single DM halos
with large occupation numbers.}

We are planning extended follow-up spectroscopy
of this newly discovered population in order to directly 
measure its real space clustering and to fully investigate its nature.
Further important steps will be the evaluation of the AGN fraction 
among $z>2$ red galaxies and their morphology. 
Work on these aspects is currently ongoing  for the Chandra Deep Field
South (CDFS/GOODS), using deep Chandra, VLT/ISAAC and ACS imaging data. 
Future SIRTF observations, planned both for HDF-S and for CDFS/GOODS, are
expected to provide constraints on the mass of $z>2$ red galaxies.

\acknowledgments

We thank ESO for the assistance in obtaining this high-quality dataset.
We thank the Lorentz Center (Leiden)
for the warm hospitality during the FIRES team meeting.
We had stimulating discussions with Alvio Renzini and Hojun Mo
on the subject of this paper. The referee, Masami Ouchi, is thanked for the
constructive report that helped improving the paper.
This work was supported by the European Community
Research and Training Network The Physics of the intergalactic medium.
\mbox{P.G.v.D.} acknowledges support by NASA through the SIRTF Fellowship program.

\onecolumn
\begin{deluxetable}{ccccrrrrr}
\tabletypesize{\scriptsize}
\tablecaption{Summary of clustering measurements in the HDF-S FIRES survey. \label{tab:all}}
\tablewidth{0pt}
\tablehead{
\multicolumn{3}{c}{Selection} & \colhead{} &  \multicolumn{5}{c}{Measurements}\\
\cline{1-3}  \cline{5-9} \\ 
\colhead{Magnitude} & \colhead{Color}   & \colhead{Phot. Redshift}   &
\colhead{} & \colhead{N} &
\colhead{$A\pm dA$\tablenotemark{a}}  & \colhead{$r_0\pm dr_0$\tablenotemark{b}} & \colhead{$z_{eff}$} &
\colhead{$\sigma_z$\tablenotemark{c}}
}
\startdata
$K<21.5$ & -- & -- &  & 125 & $4.8\pm4.3$ & $3.1^{+1.2}_{-2.2}$ & 0.7 & 0.5\\ 
$K<22.0$ & -- & -- &  & 168 & $7.3\pm3.2$ & $4.2^{+1.0}_{-1.1}$ & 0.9 & 0.8\\ 
$K<22.5$ & -- & -- &  & 217 & $4.9\pm2.5$ & $3.6^{+0.9}_{-1.2}$ & 1.0 & 0.8\\ 
$K<23.0$ & -- & -- &  & 290 & $5.1\pm1.9$ & $4.0^{+0.7}_{-0.9}$ & 1.2 & 1.0\\ 
$K<23.5$ & -- & -- &  & 362 & $4.4\pm1.5$ & $3.8^{+0.7}_{-0.8}$ & 1.1 & 1.0\\ 
$K<24.0$ & -- & -- &  & 435 & $4.1\pm1.2$ & $3.7^{+0.6}_{-0.7}$ & 1.3 & 1.1\\ 
\tableline
$K<24.0$ & -- & $0.5<z<1.0$ & & 60 & $-1.3\pm8.2$ & $0.0^{+2.7}_{-0.0}$ & -- & -- \\ 
$K<24.0$ & -- & $1.0<z<1.5$ & & 39 & $18.6\pm12.4$ & $4.8^{+1.6}_{-2.2}$ & -- & -- \\ 
$K<24.0$ & -- & $1.5<z<2.0$ & & 30 & $-0.3\pm20.5$ & $0.0^{+4.8}_{-0.0}$ & -- & -- \\ 
$K<24.0$ & -- & $2.0<z<2.5$ & & 43 & $-1.5\pm11.7$ & $0.0^{+3.1}_{-0.0}$ & -- & -- \\ 
$K<24.0$ & -- & $2.5<z<3.0$ & & 28 & $45.7\pm19.6$ & $6.7^{+1.5}_{-1.8}$ & -- & -- \\ 
$K<24.0$ & -- & $3.0<z<3.5$ & & 30 & $51.4\pm16.6$ & $6.5^{+1.1}_{-1.2}$ & -- & -- \\ 
$K<24.0$ & -- & $3.5<z<4.5$ & & 25 & $18.5\pm23.6$ & $4.2^{+2.5}_{-4.2}$ & -- & -- \\ 
$K<24.0$ & -- & $2.5<z<3.5$ & & 58 & $36.4\pm8.4$ & $6.5^{+0.8}_{-0.9}$ & -- & -- \\ 
$K<24.0$ & -- & $2.0<z<4.0$ & & 115 & $17.1\pm4.6$ & $5.5^{+0.8}_{-0.8}$ & -- & -- \\ 
\tableline
$K<24.0$ & $I-K>1.75$ & -- & & 345 & $5.7\pm1.5$ & $4.4^{+0.6}_{-0.7}$ & 1.2 & 1.0 \\ 
$K<24.0$ & $I-K>2.00$ & -- & & 287 & $8.4\pm1.8$ & $5.8^{+0.4}_{-0.7}$ & 1.5 & 0.9 \\ 
$K<24.0$ & $I-K>2.25$ & -- & & 224 & $8.7\pm2.3$ & $6.0^{+0.8}_{-0.9}$ & 1.7 & 0.8 \\ 
$K<24.0$ & $I-K>2.50$ & -- & & 155 & $6.5\pm3.3$ & $5.2^{+1.4}_{-1.7}$ & 1.8 & 0.9 \\ 
$K<24.0$ & $I-K>2.75$ & -- & & 102 & $4.5\pm5.0$ & $4.2^{+2.1}_{-4.2}$ & 1.7 & 0.8 \\ 
$K<24.0$ & $I-K>3.00$ & -- & & 69 & $-0.8\pm7.3$ & $0.0^{+4.8}_{-0.0}$ & 1.7 & 0.7 \\ 
$K<24.0$ & $I-K>3.25$ & -- & & 53 & $8.3\pm9.4$ & $5.3^{+2.8}_{-5.3}$ & 1.7 & 0.7 \\ 
$K<24.0$ & $I-K>3.50$ & -- & & 39 & $12.8\pm12.6$ & $6.2^{+2.9}_{-5.5}$ & 1.7 & 0.6 \\ 
$K<24.0$ & $I-K>3.75$ & -- & & 26 & $38.7\pm23.8$ & $12.2^{+3.7}_{-5.0}$ & 2.1 & 0.6 \\ 
$K<24.0$ & $I-K>4.00$ & -- & & 18 & $62.3\pm39.3$ & $16.9^{+5.4}_{-7.1}$ & 2.0 & 0.6 \\ 
$K<24.0$ & $I-K>3.50$ & $0.8<z<2.0$ & & 23 & $56.3\pm20.9$ & $9.7^{+2.0}_{-2.0}$ & 1.3 & 0.2 \\ 
\tableline
$K<24.0$ & $J-K>1.10$ & -- & & 369 & $4.1\pm1.4$ & $3.9^{+0.7}_{-0.9}$ & 1.5 & 1.0 \\ 
$K<24.0$ & $J-K>1.30$ & -- & & 254 & $5.8\pm2.1$ & $5.1^{+1.0}_{-1.1}$ & 2.0 & 1.1 \\ 
$K<24.0$ & $J-K>1.50$ & -- & & 158 & $5.8\pm3.3$ & $5.0^{+1.4}_{-1.9}$ & 2.5 & 1.0 \\ 
$K<24.0$ & $J-K>1.70$ & -- & & 99 & $11.7\pm5.4$ & $6.9^{+1.6}_{-2.0}$ & 2.8 & 1.0 \\ 
$K<24.0$ & $J-K>1.90$ & -- & & 59 & $29.3\pm8.9$ & $10.3^{+1.6}_{-1.9}$ & 3.1 & 0.6 \\ 
$K<24.0$ & $J-K>2.10$ & -- & & 36 & $30.5\pm14.6$ & $9.7^{+2.4}_{-2.9}$ & 3.2 & 0.6 \\ 
$K<24.0$ & $J-K>2.30$ & -- & & 24 & $59.4\pm24.4$ & $14.5^{+3.1}_{-3.7}$ & 3.4 & 0.7 \\ 
\tableline
$K<24.0$ & $J-K<1.7$ & $2.0<z<4.0$ & & 66 & $7.6\pm7.4$ & $3.5^{+1.7}_{-3.0}$ & 2.5 & 0.4 \\ 
$K<24.0$ & $J-K>1.7$ & $2.0<z<4.0$ & & 49 & $39.1\pm10.2$ & $8.3^{+1.2}_{-1.2}$ & 3.1 & 0.4 \\ 
$K<24.0$ & $J-K>1.7$ & $2.7<z<3.3$ & & 20 & $77.5\pm27.7$ & $8.2^{+1.5}_{-1.5}$ & 3.1 & 0.1 \\ 
\tableline
\enddata

\tablenotetext{a}{
In units of $10^{-4}$, amplitudes at 1 degree.
}
\tablenotetext{b}{
In units of comoving \h1 Mpc.
}
\tablenotetext{c}{
Standard deviation of the photometric redshift distribution.
}

\end{deluxetable}

\end{document}